\documentclass[aps,twocolumn,showpacs]{revtex4-1}
\usepackage{times}
\usepackage{amsmath}
\usepackage{amsfonts}
\usepackage{amssymb}
\usepackage{amsthm}
\usepackage{bm}
\usepackage{latexsym}
\usepackage{graphicx}
\usepackage{color}
\usepackage{url}
\usepackage{subfigure}

\newcommand{\mbc}{M_\mathrm{BC}}

\usepackage{hyperref}
\hypersetup{
 pdftitle={},%
 pdfauthor={},%
 pdfsubject={},%
 pdfkeywords={},%
 pdfstartview={},%
 bookmarksopen=true, breaklinks=true, debug=true,
 colorlinks=true, linkcolor=blue, citecolor=red, urlcolor=blue,
 hyperfigures=true
}
\def \b{{\cal B}}
\usepackage{relsize}
\begin{document}

\title{\boldmath Search for $D^0\to\gamma\gamma$ and improved measurement of the branching fraction for $D^0 \to \pi^0\pi^0$}
  
%-------- INSERT HERE ------------
\author{
  \begin{small}
    \begin{center}
      M.~Ablikim$^{1}$, M.~N.~Achasov$^{9,a}$, X.~C.~Ai$^{1}$,
      O.~Albayrak$^{5}$, M.~Albrecht$^{4}$, D.~J.~Ambrose$^{44}$,
      A.~Amoroso$^{48A,48C}$, F.~F.~An$^{1}$, Q.~An$^{45}$,
      J.~Z.~Bai$^{1}$
% ...
      \\
      \vspace{0.2cm}
      (BESIII Collaboration)\\
      \vspace{0.2cm} {\it
        $^{1}$ Institute of High Energy Physics, Beijing 100049, People's Republic of China\\
% ...
$^{4}$ Bochum Ruhr-University, D-44780 Bochum, Germany\\
        $^{5}$ Carnegie Mellon University, Pittsburgh, Pennsylvania 15213, USA\\
% ...
$^{9}$ G.I. Budker Institute of Nuclear Physics SB RAS (BINP), Novosibirsk 630090, Russia\\
% ...
$^{44}$ University of Rochester, Rochester, New York 14627, USA\\
        $^{45}$ University of Science and Technology of China, Hefei 230026, People's Republic of China\\
% ...
$^{48}$ (A)University of Turin, I-10125, Turin, Italy; (B)University of Eastern Piedmont, I-15121, Alessandria, Italy; (C)INFN, I-10125, Turin, Italy\\
% ...
$^{a}$ Also at the Novosibirsk State University, Novosibirsk, 630090, Russia\\
% ...
      }\end{center}
    \vspace{0.4cm}
  \end{small}
}

\affiliation{}

%-------- END INSERT ------------

%\date{\today}
\vspace{0.4cm}
%\linenumbers
\begin{abstract}
Using $2.92$ fb$^{-1}$ of electron-positron annihilation data collected at $\sqrt{s} = 3.773$~GeV with the BESIII detector, 
we report the results of a search for the flavor-changing neutral
current process $D^0\to\gamma\gamma$ using 
a double-tag technique.  We find no signal and set 
an upper limit at $90\%$ confidence level
 for the branching fraction of
$\b(D^0\to\gamma\gamma) < 3.8\times10^{-6}$.  We also investigate
$D^0$-meson decay 
into two neutral pions, obtaining a branching fraction of 
$\b(D^0\to\pi^0\pi^0) =
(8.24\pm0.21(\text{stat.})\pm0.30(\text{syst.}))\times10^{-4}$, the
most precise 
measurement to date and consistent with the current world average.

\end{abstract}


\maketitle

\section{\boldmath Introduction}\label{sec:intro}

In the standard model (SM), the flavor-changing neutral current (FCNC) decay $D^0 \rightarrow \gamma
\gamma$ is strongly suppressed by the Glashow-Iliopoulos-Maiani mechanism~\cite{gim1970}. 
 The branching fraction for $D^0\rightarrow \gamma
\gamma$  from short-distance contributions, such as
an electromagnetic penguin transition, is predicted to be $3\times
10^{-11}$~\cite{greub1996,fajfer2001,burdman2002}. 
 Long-distance contributions due to a vector meson coupling to a
 photon are expected to enhance the branching fraction
to the range $(1-3)\times 10^{-8}$~\cite{fajfer2001,burdman2002}.  
These predictions are orders of magnitude beyond the reach of current experiments, but some extensions to the SM can 
enhance FCNC processes by many orders of magnitude.  For example, in
the framework of the minimal supersymmetric SM, gluino exchange can
increase the branching fraction for the $c\rightarrow u \gamma$ transition to $6\times 10^{-6}$~\cite{Prelovsek2001,bigi2010}. 

The previous experimental studies of $D^0\rightarrow \gamma \gamma$
were performed by the CLEO and {\it BABAR} experiments using data samples
collected at the  $\Upsilon(4S)$ peak~\cite{cleo,babar}.  
With an integrated luminosity of $470.5$ fb$^{-1}$,  corresponding to
more than $250$ million $D^0$ mesons based on the quoted number of reconstructed $D^0\to\pi^0\pi^0$ candidates,
its efficiency, and the measured $\b(D^0\to\pi^0\pi^0)$ in Ref.~\cite{babar},
{\it BABAR} set an upper limit at
$90\%$ confidence level (CL) on the $D^0\to\gamma\gamma$ branching fraction of
$2.2\times 10^{-6}$ which is the most stringent limit to date.

In this paper we report a search for $D^0\rightarrow \gamma \gamma$ 
using $2.92\pm0.03$ fb$^{-1}$  of 
$e^+e^-$ annihilation data collected by the BESIII
detector~\cite{bes3lumi} 
at $\sqrt{s} = 3.773$~GeV in $2010$ and $2011$.
There are about $20$ million $D^0$ mesons produced~\cite{He:2005bs}
from $\psi(3770)$ decays in this sample.
Taking advantage of the fact that $D$-meson production near the
$\psi(3770)$ resonance is solely through $D\bar{D}$, we apply a tagged
technique pioneered by the MARK III Collaboration~\cite{marck3}.
After reconstructing a hadronically decaying $\bar{D}$ in an event
(the tag), we then search for $D$-decay candidates of interest in the
remainder of the event. (Unless
otherwise noted, charge conjugate modes are implied throughout this
paper.)
 This strategy suppresses background and
provides an absolute normalization for branching fraction measurements
independent of the integrated luminosity and $D\bar{D}$ production
cross section.  
Therefore,  searches for $D^0 \to \gamma \gamma$ with BESIII at
open-charm threshold are uniquely clean and 
provide a valuable complement to studies at the $\Upsilon(4S)$. 

In addition to our primary result, we also report an improved measurement of the branching fraction for the decay $D^0 \to \pi^0 \pi^0$, which is the dominant background for $D^0  \to \gamma \gamma$. Precise measurement of the $D^0 \to \pi^0 \pi^0$ branching fraction can improve understanding of U-spin and SU(3)-flavor symmetry breaking effects in $D^0$ decays~\cite{Grossman:2012ry}, benefiting theoretical predictions of $CP$ violation in $D$ decays~\cite{Grossman:2012eb}.  

%_/_/_/_/_/_/_/_/_/_/_/_/_/_/_/_/_/_/_/_/_/_/_/_/_/_/_/_/_/_/_/_/_/_/_/_/_/_/
\section{\boldmath the BESIII detector and Monte Carlo simulations}\label{sec:samples}
 
The data used in this analysis were collected 
with the BESIII detector operating at the BEPCII Collider.
The BESIII detector, which is described in detail elsewhere~\cite{bes3det},
has a geometrical acceptance of $93\%$ of $4\pi$ and 
consists of four main components.  A small-celled, helium-based,
multilayer drift chamber (MDC) with $43$ layers provides  momentum
resolution for $1$-GeV/$c$ charged particles in a $1$-T magnetic field of $0.5\%$.
Excellent charged particle identification is achieved 
by utilizing the energy loss in the MDC ($dE/dx$).
A time-of-flight system (TOF) for additional charged particle identification is composed
of plastic scintillators.
The time resolution is $80$~ps in the barrel
and $110$~ps in the endcaps, giving $2\sigma$ $K$/$\pi$
separation for momenta up to about 
$1$~GeV/$c$.
An electromagnetic calorimeter (EMC) is constructed of $6240$ CsI (Tl)
crystals arranged in a cylindrical shape (barrel) plus two endcaps. 
For $1.0$-GeV photons, the energy resolution is $2.5\%$ in the barrel
and $5\%$ in the endcaps.
Finally, a muon chamber system (MUC) is constructed of 
resistive plate chambers.
These are interleaved with the flux-return iron of the superconducting magnet. 

Monte Carlo (MC) simulations are used for efficiency and background determinations.  Events
are generated with {\sc kkmc}~\cite{kkmc}, which incorporates initial-state radiation and the spread of the
BEPCII beam energy. The generated particles are subsequently passed to {\sc evtgen}~\cite{evtgen}, which simulates particle decays
based on known branching fractions~\cite{pdg}.
To realistically mimic our data, we produce a generic MC sample including $e^+e^-\to\psi(3770)\to D\bar{D}$, continuum hadron production ($e^+e^-\to\gamma^*\to q\bar{q}$, with $q = u, d$ or $s$), radiative returns to the lower $c\bar{c}$ resonances
($e^+e^-\to\gamma_{ISR} (\psi(3686)$ or $J/\psi$)), $e^+e^- \to \tau^+\tau^-$, and  the doubly-radiative Bhabha process $e^+ e^- \to e^+e^-\gamma \gamma $.  The last component is generated with {\sc Babayaga}~\cite{babayaga}.  
We also generate  a signal MC sample consisting of $e^+e^-\to\psi(3770) \to D^0\bar{D}^0$ events in which the $D^0$ or the $\bar{D}^0$ decays into a hadronic tag mode or $\gamma \gamma$, while the other $\bar{D}^0$ or $D^0$ decays without restriction.  For all MC samples, generated events are processed with {\sc geant4}~\cite{geant} to simulate the BESIII detector response.\\

%_/_/_/_/_/_/_/_/_/_/_/_/_/_/_/_/_/_/_/_/_/_/_/_/_/_/_/_/_/_/_/_/_/_/_/_/_/_/
\section{\boldmath $D^0 \to \gamma \gamma$ analysis with  Double-tag method}\label{sec:doubletag}

The $\psi(3770)$ resonance is below the threshold for $D\bar{D} \pi$ production, so the events from $e^+e^- \to \psi(3770) \to D\bar{D}$
have $D$ mesons with energies equal to the beam energy
($E_{\text{beam}}$) and known momentum. 
Thus, to identify 
$\bar{D}^0$ candidate, 
we define the two variables 
$\Delta E$ and $\mbc$,
the beam-constrained mass:
\begin{equation}\label{eq:variables}
\begin{split}
 & \Delta E  \equiv \sum_i E_i - E_{\text{beam}},  \\
 & \mbc  \equiv \sqrt{ E^2_{\text{beam}} - | \sum_i \vec{p}_i|^2} , \nonumber
\end{split}
 \end{equation}
where $E_i$ and $\vec{p}_i$ are the energies and momenta of the $\bar{D}^0$ decay
products in the center-of-mass system of the $\psi(3770)$.
For true $\bar{D}^0$ candidates, $\Delta E$  will be 
consistent with zero, and $\mbc$ will be consistent with the $\bar{D}^0$ mass.

Single tag (ST) candidate events are selected by reconstructing a $\bar{D}^0$
in one of the following five hadronic final states:
 $\bar{D}^0\to K^{+}\pi^{-}$, $K^{+}\pi^{-}\pi^0$, $K^{+}\pi^{-}\pi^{+}\pi^{-}$,
$K^{+}\pi^{-}\pi^{+}\pi^{-}\pi^0$, and $K^{+}\pi^{-}\pi^0\pi^0$,
constituting approximately 37\% of all $\bar{D}^0$ decays~\cite{pdg}. 
The resolution of $\mbc^{\text{tag}}$ is about 2~MeV/$c^2$, dominated by 
the beam-energy spread.  The $\Delta E^{\text{tag}}$ resolutions are 
about $10$~MeV and $15$~MeV 
for final states consisting entirely of charged tracks and for those
including a $\pi^0$, respectively.
We search for $D^0 \to \gamma \gamma$ decays in these tagged events,
 thereby highly suppressing backgrounds from QED continuum processes,
potential $\psi(3770) \to \text{non}$-$D\bar{D}$ decays, as well as
$D^+D^-$ decays.
The fraction of double tag (DT) events, in which the $D^0$ is
reconstructed as  $D^0 \to \gamma \gamma$, determines the absolute
 branching fraction for the signal mode, 
\[
 \b(D^0\to\gamma\gamma) = \frac{N_{\text{tag},\gamma\gamma}}{\sum_{i}{N_{\text{tag}}^{i}\cdot(\epsilon^i_{\text{tag}, \gamma\gamma}/\epsilon^i_{\text{tag}})}}. \\
\]
\noindent In this expression  $i$ runs over each of the five tag
modes, $N_{\text{tag}}$ and $\epsilon_{\text{tag}}$ are the ST yield
and reconstruction efficiency,
 and  $N_{\text{tag},\gamma\gamma}$ and
 $\epsilon_{\text{tag},\gamma\gamma}$ are the yield and efficiency 
for the DT combination of a hadronic tag 
and a $D^0 \to \gamma \gamma$ decay.  
 
\subsection{\boldmath Single-tag selection and yields}\label{subsec:stag} 

 For each tag mode, $\bar{D}^0$ candidates are reconstructed from all
possible combinations of final-state particles, according to the following selection criteria. 
 Momenta and impact parameters of charged tracks are measured by the
 MDC. 
Charged tracks are required to satisfy $|\text{cos}\theta| <0.93$, 
where $\theta$ is the polar angle with 
 respect to the 
direction of the positron beam,
and to have a closest approach to the 
interaction point
 within $\pm 10$~cm along the beam direction and within $1$~cm in the
 plane 
perpendicular to the beam.
 Discrimination of charged pions from kaons is achieved by combining
 information 
about the normalized energy deposition ($dE/dx$) in the MDC with the
flight-time 
measurement from the TOF. 
 For a positive identification, the probability of the $\pi$($K$)
 hypothesis is required to
 be larger than that of the $K$($\pi$) hypothesis. 
 
Electromagnetic showers are reconstructed from clusters of energy
deposits in the EMC crystals and are required 
to be inconsistent with deposition by
charged tracks~\cite{y_cp}.
The energy deposited in nearby TOF counters is included to improve the
reconstruction efficiency and energy resolution. 
 The shower energies are required to be greater than $25$ MeV for the
 barrel region ($|\text{cos}\theta|<0.80$) and
greater than $50$ MeV for the endcaps ($0.84 < |\text{cos}\theta| < 0.92$). 
 Showers in the angular range between the barrel and endcaps are
 poorly reconstructed and 
excluded from the analysis. Cluster-timing requirements are used to
suppress 
electronic noise and energy deposits 
  unrelated to the event.  For any tag mode with a $\pi^0$ in the
  final state,  photon pairs are used to reconstruct $\pi^0$
  candidates if the invariant mass satisfies 
$(115<m_{\gamma \gamma}<150)$ MeV/$c^2$.  
To improve resolution and reduce background, we constrain the
invariant mass of each 
photon pair to the nominal $\pi^0$ mass. 

\begin{figure}[htb]
\begin{center}
\includegraphics[keepaspectratio=true,width=3.4in,angle=0]{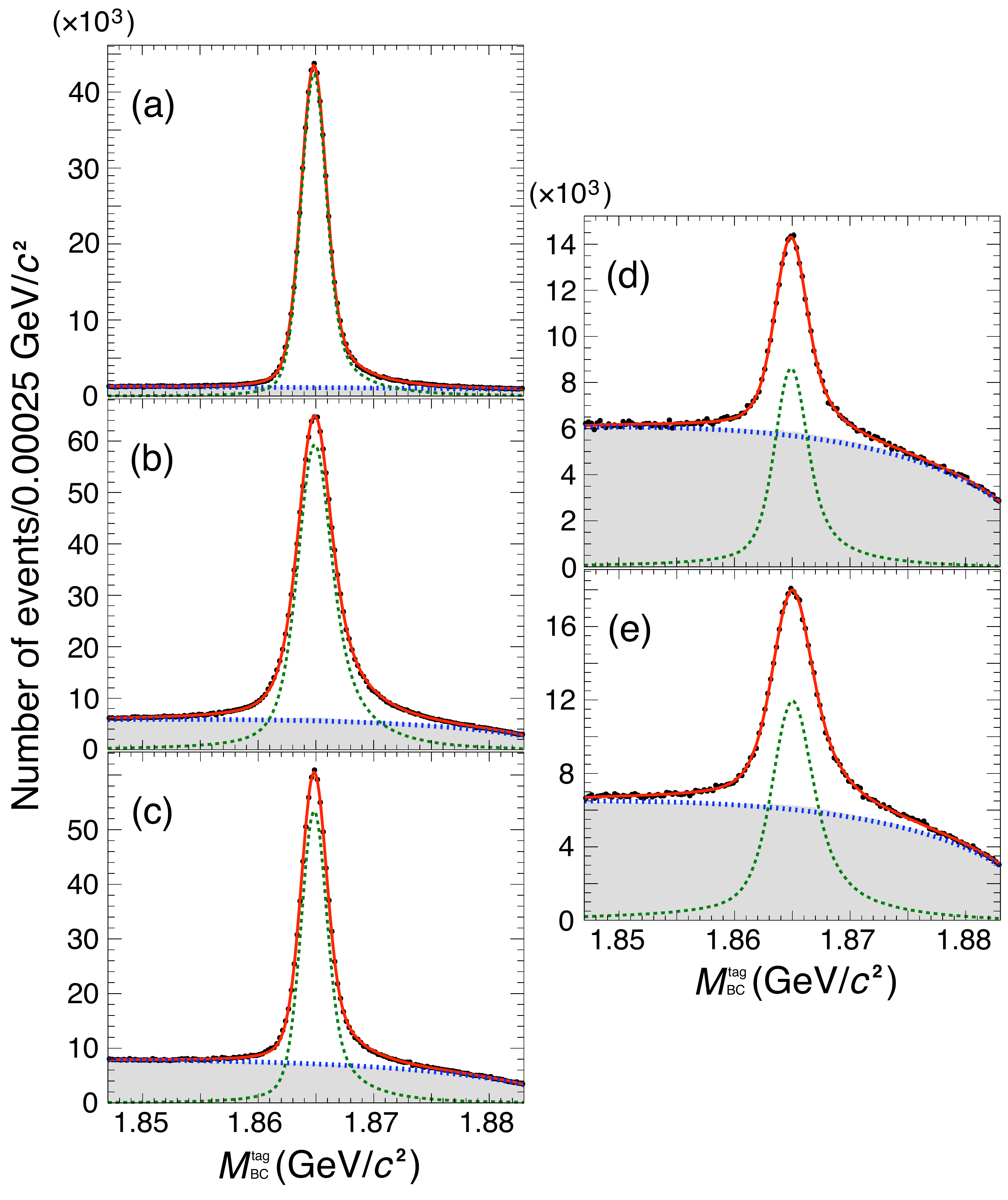}
\caption{Fits (solid line) to the $\mbc^{\text{tag}}$ distributions in data (points) for the five $\bar{D}^0$ tag modes: 
(a) $ K^{+}\pi^{-}$, (b) $K^{+}\pi^{-}\pi^0$,
(c) $K^{+}\pi^{-}\pi^{+}\pi^{-}$,
(d) $K^{+}\pi^{-}\pi^{+}\pi^{-}\pi^0$, and
(e) $K^{+}\pi^{-}\pi^0\pi^0$.
The gray shaded histograms are arbitrarily scaled generic MC backgrounds.
}\label{fig:stdtfit1}
\end{center}
\end{figure}

For ST modes,  we accept $\bar{D}^0$ candidates that satisfy 
the requirements $1.847<\mbc^{\text{tag}}<1.883$~GeV/$c^2$ and 
$|\Delta E^{\text{tag}}|<0.1$~GeV.  
 In events with multiple tag candidates, the one candidate per mode
 with reconstructed energy 
closest to the beam energy is chosen~\cite{He:2005bs}.
We extract the ST yield for each tag mode and the combined yields of
all five modes from fits to $\mbc^{\text{tag}}$ distributions in the
samples described above. 
The signal shape is derived from the MC simulation
which includes the effects of 
beam-energy smearing, initial-state radiation, the $\psi(3770)$ line
shape, and detector resolution.
We then convolute the line shape with a Gaussian
to compensate for a difference
in resolution between data and our MC simulation.
Mean and width of the convoluted Gaussian, along with the
overall normalization, are left free in our nominal fitting procedure.
The background is described by an ARGUS function~\cite{argusBKG}, 
 which models combinatorial contributions. 
In the fit, we leave free all parameters of  the background function, except
its endpoint which is fixed at $1.8865$~GeV/$c^2$.
Figure~\ref{fig:stdtfit1}
 shows the fits to our tag-candidate samples.  Tag yields, given in
 Table~\ref{tab:dtefftest}, are obtained by subtracting the
fitted background estimates from the overall fits
in data within the narrow signal window $\mbc^{\text{tag}}$  
($1.858<\mbc^{\text{tag}}<1.874$~GeV$/c^2$).
The total number of tags reconstructed in our data is approximately
$2.8$ million. 
 Also shown in Table~\ref{tab:dtefftest} are the tagging efficiencies
 obtained by fitting generic 
MC $\mbc^{\text{tag}}$ distributions with the same procedure used on
data.  
These ST and DT efficiencies include the $\pi^0 \to \gamma \gamma$ branching fraction. 
  
 \begin{table}[htb]
\caption{ Single-tag efficiencies ($\epsilon^i_{\text{tag}}$),  tag
  yields ($N^i_{\text{tag}}$) in data,  double-tag efficiencies
  ($\epsilon^i_{\text{tag}, \gamma \gamma}$) and their statistical
  uncertainties. 
Efficiencies are determined based on MC simulations.
}
\label{tab:dtefftest}
\begin{center}
\begin{tabular}{c| c r c r r c c  }
\hline
modes         & $\epsilon^i_{\text{tag}}$ (\%) & \multicolumn{3}{c}{$N^i_{\text{tag}}$} & \multicolumn{3}{c}{$\epsilon^i_{\text{tag}, \gamma \gamma}$ (\%)}    \\
\hline
\multicolumn{1}{l|}{$K^+\pi^-$}            & $66.12\pm0.04$ &   $551800$ & $\pm$ &   $936$ & $44.8$ &  $\pm$ &  $0.4$ \\
\multicolumn{1}{l|}{$K^+\pi^-\pi^0$}    & $35.06\pm0.02$ & $1097113$ & $\pm$ & $1386$ & $24.5$ & $\pm$ &  $0.1$ \\
\multicolumn{1}{l|}{$K^{+}\pi^{-}\pi^{+}\pi^{-}$}          & $39.70\pm0.03$ &   $734825$ & $\pm$ & $1170$ & $24.7$ & $\pm$ &  $0.2$ \\
\multicolumn{1}{l|}{$K^{+}\pi^{-}\pi^{+}\pi^{-}\pi^0$}  & $15.32\pm0.04$ &   $155899$ & $\pm$ &   $872$ & $9.6$ & $\pm$ &  $0.1$  \\
\multicolumn{1}{l|}{$K^{+}\pi^{-}\pi^0\pi^0$} &  $15.23\pm0.04$ &  $268832$ &
                                                                    $\pm$ &   $976$ & $8.9$ & $\pm$ &  $0.1$ \\ \hline
\multicolumn{1}{l|}{All Tags}          &                               &  $2808469$   & $\pm$ &   $2425$  &      &     &   \\  
\hline
\end{tabular}
\end{center}
\end{table}
 
\subsection{\boldmath Double-tag selection and yield}\label{subsec:dtag} 
 
 We select DT candidates by reconstructing $D^0 \to \gamma \gamma$ from the
 two most energetic photon candidates that are not used in
 reconstructing the tag mode.
The selection criteria for these photons are the same as the ones used
on the tag side, except that we require
$0.86<|\cos{\theta}|<0.92$ for endcap showers
to remove photons landing near the transition region.
We require $|\Delta E^{\text{tag}}|<0.10$~GeV
 ($1.858<\mbc^{\text{tag}}<1.874$~GeV/$c^2$) and 
$|\Delta E^{\gamma \gamma}|<0.25$~GeV 
($\mbc^{\gamma\gamma}>1.85$~GeV/$c^2$)
to the tag $\bar{D}^0$ candidate and the signal $D^0$ candidate, respectively.  
If there are multiple DT candidates, we choose the combination for which the average of $\mbc^{\text{tag}}$ and 
$\mbc^{\gamma\gamma}$ ($\bar{M}_{\mathrm{BC}} \equiv ( \mbc^{\text{tag}}+  \mbc^{\gamma\gamma})/2$) is  closest to the known $D^0$ mass~\cite{He:2005bs}. 
  
\begin{figure}[htb]
\begin{center}
\includegraphics[keepaspectratio=true,width=3.2in,angle=0]{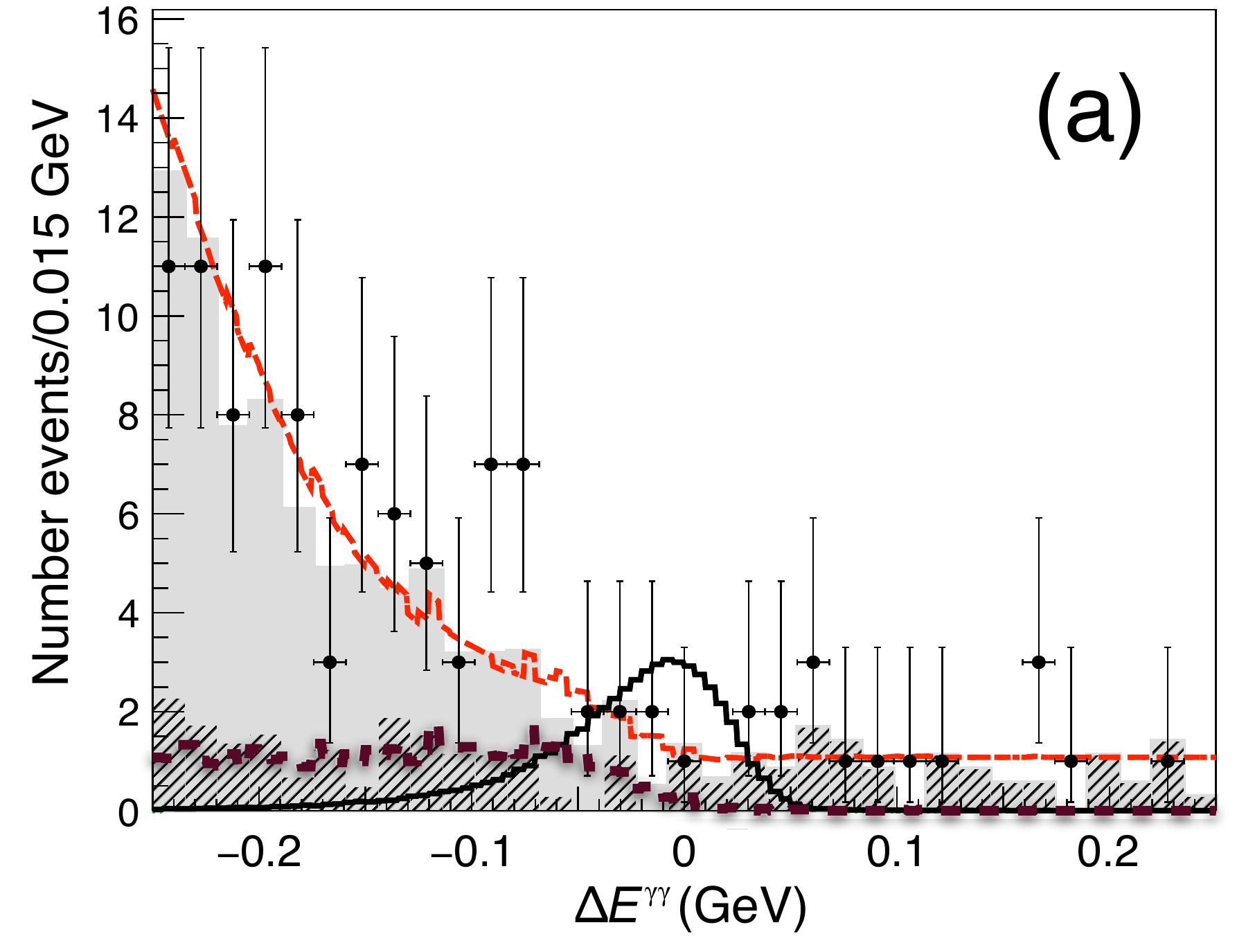}\\
\vspace{0.4cm}
\includegraphics[keepaspectratio=true,width=3.2in,angle=0]{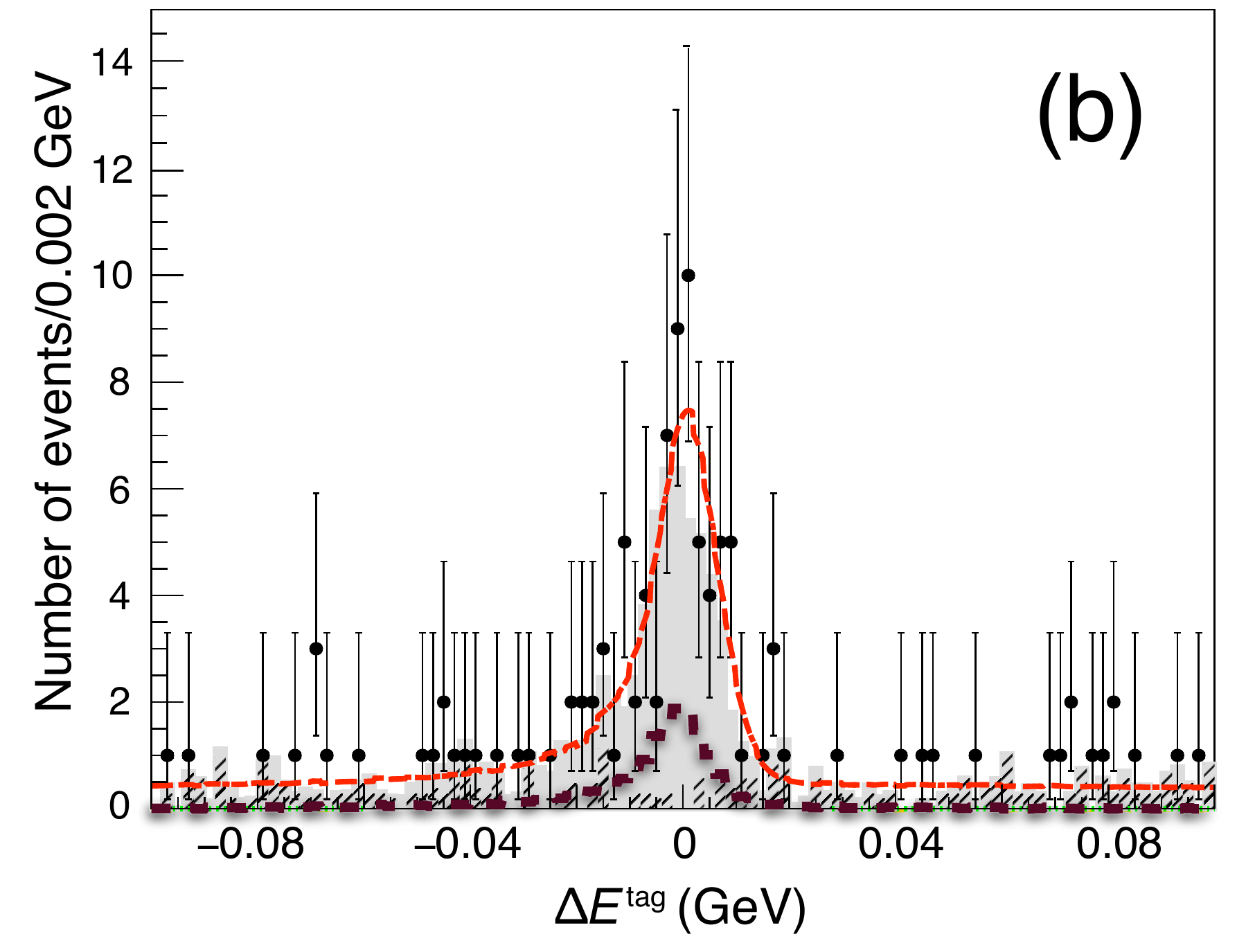} 
\caption{Fit to the DT sample in data (points), projected onto $\Delta E^{\gamma\gamma}$ (a) and
$\Delta E^{\text{tag}}$ (b).
The dashed lines show the overall fits, while the
dotted histograms represent the estimated background
contribution from $D^0\to\pi^0\pi^0$.
The solid line superimposed on the $\Delta E^{\gamma\gamma}$  projection indicates
the expected signal for $\b(D^0\to\gamma\gamma) = 10\times10^{-6}$.
Also overlaid are the overall MC-estimated backgrounds (gray shaded histograms) and the background
component from non-$D\bar{D}$ processes (diagonally hatched histograms).
}\label{fig:dtggdatafit}
\end{center}
\end{figure}

For any DT including  $\bar{D}^0 \to K^+ \pi^-$, the dominant background is from the doubly-radiative Bhabha 
QED process $e^+e^- \to e^+ e^- \gamma \gamma$, which has a large production cross-section. 
 To remove this background, we require the angle between the direction of the photon
candidates and any charged tracks to be greater than 10 degrees.
This requirement eliminates 
$93\%$ of the QED background.  For all tag modes, the dominant peaking background  in the 
$\Delta E^{\gamma \gamma}$ signal region is from $D^0 \to \pi^0 \pi^0$. To remove this background, we implement a $\pi^0$ veto.  We reject
 events in which one of the $D^0 \to \gamma \gamma$ final-state photons can be combined with any other photon
 in the event to form a $\pi^0$.  
This requirement rejects 
$82\%$
of the $D^0 \to \pi^0 \pi^0$ background and keeps
$ 88\%$
of the signal events.  
Figure~\ref{fig:dtggdatafit} shows the distributions of $\Delta E^{\gamma\gamma}$ (top)
and $\Delta E^{\text{tag}}$ (bottom)  after the above selection criteria are applied, overlaid with the MC background estimate.

While we can suppress most of the background with the DT method,
there remain residual contributions from continuum processes, 
primarily doubly-radiative Bhabha events for $K\pi$ tags and $e^+e^-
\to q\bar{q}$ for other modes.
In order to correctly estimate their sizes, we take a  data-driven
approach by  performing an unbinned  maximum likelihood
fit to the two-dimensional distribution of   $\Delta E^{\gamma\gamma}$
versus $\Delta E^{\text{tag}}$. 
We use $\Delta E^{\gamma\gamma}$  distributions rather than
$\mbc^{\gamma\gamma}$ distributions as the background from
non-$D\bar{D}$ decays is more easily addressed in the fit.
Also, the background from
$D^0\to\pi^0\pi^0$ peaks in $\mbc^{\gamma\gamma}$ at the same place
as the signal does, whereas it is shifted in $\Delta E^{\gamma\gamma}$.
The fitting ranges are $|\Delta E^{\gamma\gamma}|<0.25$~GeV and 
$|\Delta E^{\text{tag}}|<0.1$~GeV.
These wide ranges are chosen to have adequate statistics of the
continuum backgrounds in our fit.
The $\Delta E^{\gamma\gamma}$ resolution is $25$~MeV, as determined with signal MC. 
For the signal and the $D^0 \to \pi^0 \pi^0$ background, we extract probability density functions (PDFs) from  MC,
where the number of $D^0\to\pi^0\pi^0$ background events is fixed
to the result of the data-driven method described in Sec.~\ref{sec:dt00}.
For the background from continuum processes,
we include a flat component in two dimensions, allowing the
normalization to float.
The contribution from $D^+D^-$ decays is completely negligible.
We model the background from other $D^0\bar{D}^0$ decays with a pair of functions.  In the $\Delta E^{\text{tag}}$ 
dimension  we use a Crystal Ball Line function (CBL)~\cite{cblshape} plus a 
Gaussian, and in the  $\Delta E^{\gamma\gamma}$ dimension, we use a second-order exponential polynomial: 
\[
 Y(\Delta E^{\gamma\gamma}) = N\times e^{- (c_1\cdot\Delta E^{\gamma\gamma} + c_2\cdot(\Delta E^{\gamma\gamma})^2)}.
\]
In our nominal fitting procedure, we fix the following parameters
based on MC: the power-law tail parameters of the CBL,
the coefficients ($c_1$ and $c_2$) of the above exponential
polynomial, and the mean and the width of the Gaussian function.
The normalization for the background from all other
$D^0\bar{D}^0$ decays is left free in the fit, 
as are the mean and width of the CBL and
the ratio of the areas of the CBL and Gaussian functions.
Table~\ref{tab:dtefftest} lists the DT
signal-reconstruction efficiencies for each of the five tag modes.

As a test to validate the fitting procedure,
we fit to $10,000$ sets of pseudo-data (toy MC samples) generated
by randomly distributing points based on our generic MC samples
while taking into account the Poisson distribution
with input $D^0 \to \gamma \gamma$ branching
fractions of 
$(0,5,10)\times10^{-6}$. The average branching fractions measured
with these samples are 
$(0.3\pm1.2,5.0\pm2.4,10.0\pm3.1)\times10^{-6}$,
respectively, where the quoted uncertainties are the root-mean-squares
of the distributions.

Figure~\ref{fig:dtggdatafit} shows projections of
the fit to the DT data sample onto $\Delta E^{\gamma\gamma}$ (top)
and $\Delta E^{\text{tag}}$ (bottom).
We also overlay background distributions predicted by the MC simulations.
The fit yields  $N_{\text{tag},\gamma\gamma}=(-1.0^{+3.7}_{-2.3})$,
demonstrating that there is no signal for $D^0 \to \gamma \gamma$ in our data.
This corresponds to $\b(D^0\to\gamma\gamma) =  (-0.6^{+2.0}_{-1.3})\times10^{-6}$
where the uncertainties are statistical only.

\section{\boldmath Size of  $D^0\to\pi^0\pi^0$ background}\label{sec:dt00}

To estimate the contribution of background from $D^0\to\pi^0\pi^0$ events to our selection, we make a second 
DT measurement with the same sample used in searching for $D^0 \to \gamma \gamma$.  Within these tagged events, we reconstruct 
$D^0 \to \pi^0\pi^0$ with the $\pi^0$ candidates that are not used in reconstructing the tag modes.  
The selection criteria for these $\pi^0$ candidates are the same as those used in reconstructing the tags.  We select the pair of 
$\pi^0$s that gives the smallest $|\Delta E^{\pi^0\pi^0}|$ and extract
the DT yield by fitting to $\mbc^{\pi^0\pi^0}$, while requiring
$-0.070<\Delta E^{\pi^0\pi^0}<+0.075$~GeV. 
In this fit, a double-Gaussian function is used to represent the
$\mbc^{\pi^0\pi^0}$
shape for the $D^0\to \pi^0\pi^0$ decays, while
the $D^0\bar{D}^0$ MC shape describes the background.

\begin{figure}[htpb]
\begin{center}
\includegraphics[keepaspectratio=true,width=3.3in,angle=0]{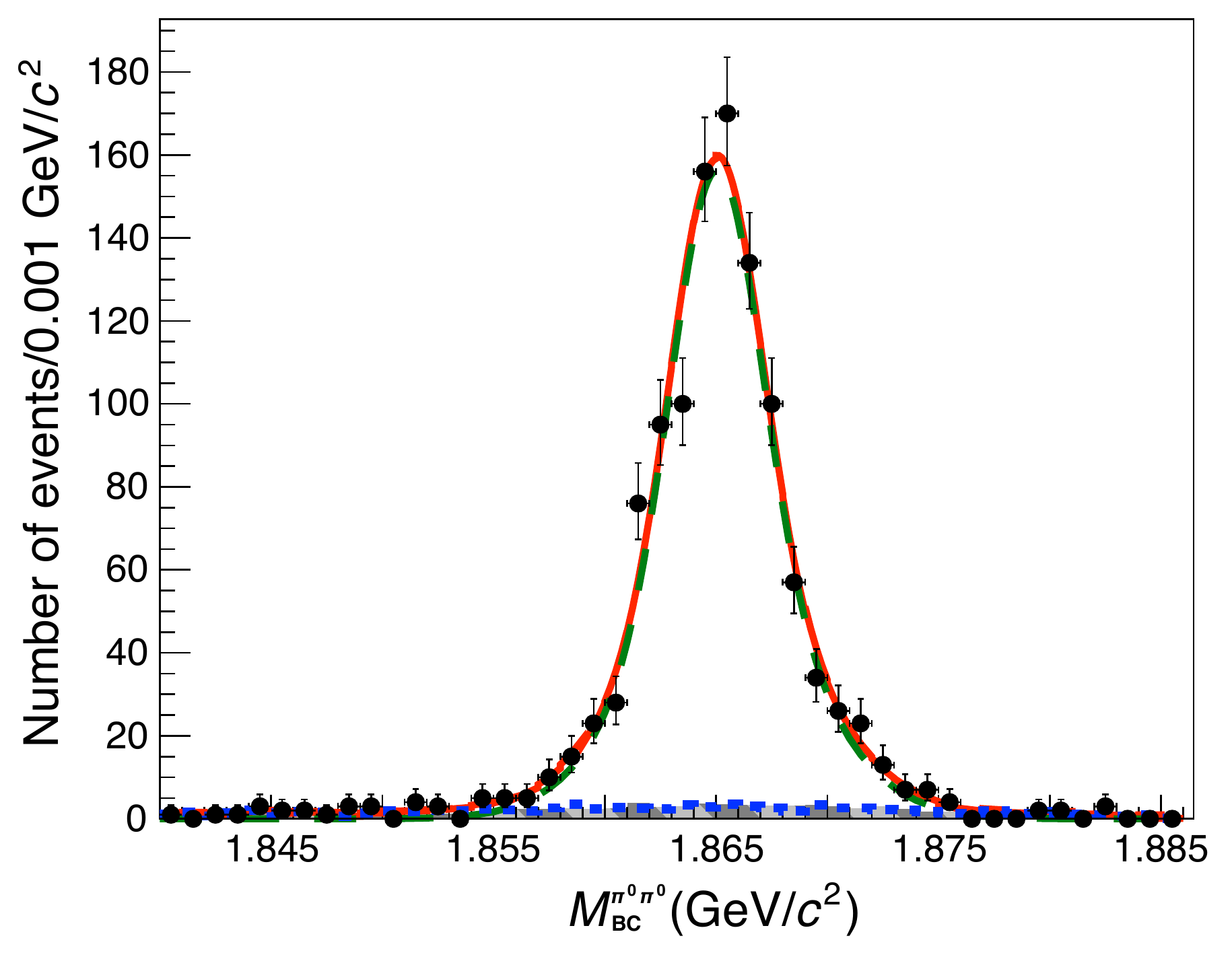}
\caption{Fit to  the $\mbc^{\pi^0\pi^0}$ distribution in data (points) for $D^0\to\pi^0\pi^0$ DT candidates.
The solid line is the total fitted result, while the dotted and
dashed lines are the background and signal components of the fit, respectively.
The diagonally shaded histogram is the background determined with MC.
}\label{fig:dt00dtfit}
\end{center}
\end{figure}

Figure~\ref{fig:dt00dtfit} shows the fit to the $\mbc^{\pi^0\pi^0}$
distribution in $1.840<\mbc^{\pi^0\pi^0}<1.886$~GeV/$c^2$,
which yields
 $N^{\text{obs}}_{\pi^0\pi^0} = 1036\pm35$  events for $D^0\to \pi^0\pi^0$.  Thus the yield in our data sample of 
 $D^0 \to \pi^0\pi^0$ with a $\bar{D}^0$ decaying into one of the five tag modes is 
 $N^{\text{produced}}_{\pi^0\pi^0} =  N^{\text{obs}}_{\pi^0\pi^0}
 /\epsilon^{\pi^0\pi^0}_{\text{DT}}$, 
where $\epsilon^{\pi^0\pi^0}_{\text{DT}}=6.08\%$ is the DT efficiency for $D^0 \to \pi^0\pi^0$ as determined with MC. 
 The expected $\pi^0\pi^0$ contribution to our $\gamma\gamma$ candidates
can be then obtained as 
\[
 N^{\text{expected}}_{\pi^0\pi^0} =N^{\text{produced}}_{\pi^0\pi^0} \times   \epsilon^{\gamma\gamma}_{\pi^0\pi^0} =N^{\text{obs}}_{\pi^0\pi^0} \frac{ \epsilon^{\gamma\gamma}_{\pi^0\pi^0}}{\epsilon^{\pi^0\pi^0}_{\text{DT}} }
\]
where  $\epsilon^{\gamma\gamma}_{\pi^0\pi^0}=0.11\%$ is the efficiency for $D^0  \to\pi^0 \pi^0$ to be counted as  $D^0 \to \gamma\gamma$. 
The efficiencies  $\epsilon^{\gamma\gamma}_{\pi^0\pi^0}$ and $\epsilon^{\pi^0\pi^0}_{\text{DT}} $ include the
reconstruction efficiencies for the tag sides as well as the branching fractions, although these cancel in the ratio.

We consider the following sources of systematic uncertainty in
determining the $D^0 \to \pi^0\pi^0$ contamination: 
$\pi^0$ reconstruction ($1.5\%$), photon reconstruction ($2.0\%$),
binning of $\mbc^{\pi^0\pi^0}$ ($0.1\%$), fit range ($0.1\%$),
background shape ($0.5\%$), signal shape ($1.7\%$), and
the $\Delta E^{\pi^0\pi^0}$ requirement ($0.6\%$).
Combining statistical and systematic uncertainties, we estimate
the number of $D^0\to\pi^0\pi^0$ events among the $D^0\to\gamma\gamma$ candidates
to be $18$ events with a relative uncertainty of $4.6\%$,
spread across the $\Delta E^{\gamma\gamma}$ fit range.  

\section{\boldmath Systematic uncertainties for $D^0 \to \gamma \gamma$  analysis}\label{sec:dtgamgamsyst}

MC studies demonstrate that $D$-decay measurements based on DT-to-ST ratios benefit from cancellation of 
most of the systematic uncertainties of tag reconstruction.  The overall systematic uncertainty in our measurement 
is therefore dominated by other effects.
The systematic uncertainties that are independent of our signal-fitting procedure are that associated
with detection of the two photons, which is estimated 
by studying the reconstruction efficiency of a daughter photon from
$\pi^0$ decay in a  DT $D^0\to K_S^0\pi^0$ sample ($2.0\%$);
the signal-side $\mbc^{\gamma\gamma}$ requirement, which is estimated from the
$\Delta E^{\pi^0\pi^0}$ distribution of the DT $D^0\to\pi^0\pi^0$
sample and by observing the stability of the $\b(D^0\to\pi^0\pi^0)$
while varying the selected range of $\mbc^{\pi^0\pi^0}$ ($3.1\%$).
The systematic uncertainties in ST yields  ($1.0\%$) are 
estimated first for individual tag modes, and then combined in quadrature with 
weights based on the observed tag yields ($N^{i}_{\text{tag}}$). 
The sources for the uncertainties of ST yields
we consider are the choice of fit range, assumed signal
parametrization, and the $\mbc^{\text{tag}}$ signal window.
Combined in quadrature, these total $3.8\%$.

We also consider six possible sources of systematic effects due to our fitting procedure.
(i) Fits are redone with all possible combinations of fitting ranges:
$-(0.12$$,0.10$$,0.08)$$<$$\Delta
E^{\text{tag}}$$<$$+(0.08$$,0.10$$,0.12)$~GeV and
 $-(0.30$$,0.25$$,0.20)$$<$$\Delta
 E^{\gamma\gamma}$$<$$+(0.20$$,0.25$$,0.30)$~GeV.
(ii) The MC-based analytic form of the $D^0\bar{D}^0$ background shape
(excluding the $D^0\to\pi^0\pi^0$ contribution)
is varied by changing the input branching fractions for $D^0\to \pi^0\eta/\eta\eta/K^0_L\eta/K^0_L\pi^0$ 
by $\pm 1\sigma_{\text{PDG}}$~\cite{pdg}.
(iii) The flat non-$D\bar{D}$ background shape is replaced with a shape that is linear in the $\Delta E^{\gamma\gamma}$ dimension.
(iv) The fixed size of the background from $D^0\to\pi^0\pi^0$ is varied by
$\pm4.6\%$.
(v) The fixed shape of the background from $D^0\to\pi^0\pi^0$ is studied
 by comparing $\Delta E$ distributions of DT
events from $D^0\to\pi^0\pi^0/K_S^0\pi^0/K\pi\pi^0$
between data and MC simulations in which we intentionally ignore
the lower-energy photon from each $\pi^0$ decay to mimic
our background. 
We conclude that we do not need to assign additional systematic
uncertainty due to the assumed $D^0\to\pi^0\pi^0$ background shape in
the fit,
except to give an extra Gaussian smearing of $\sigma = 5$~MeV 
in the $\Delta E^{\text{tag}}$ dimension.
(vi) The fixed signal shape is studied 
based on the DT $D^0\to\pi^0\pi^0$ sample
in which we study distributions of its $\Delta E^{\text{tag}}$ and
$\Delta E^{\pi^0\pi^0}$ for four cases by requiring that one of the
two photons from each of the two $\pi^0$ to have at least
$0.5$, $0.6$, $0.7$, and $0.8$~GeV to mimic our signal photon energies.
From all four cases,
we find that we need an extra Gaussian smearing of $\sigma = 16$~MeV and a shift by a factor of $1.0025$
in the $\Delta E^{\gamma\gamma}$ dimension
as well as an extra smearing of $\sigma = 5$~MeV 
in the $\Delta E^{\text{tag}}$ dimension.

Table~\ref{tab:dtggsyst} summarizes 
systematic uncertainties that are independent of our fitting procedure,
as well as systematic variations that we consider to estimate
uncertainties due to the fitting procedure.
In the next section, we describe how we combine these systematic
uncertainties into our measurement.

\begin{table}[htb]
\caption{Systematic uncertainties and variations
for $D^0\to\gamma\gamma$ analysis.
}
\label{tab:dtggsyst}
\begin{center}
\scalebox{1.0}
{
\begin{tabular}{c| c }

\multicolumn{2}{c}{Uncertainties independent of fitting procedure} \\
\hline
 Source & Relative uncertainty $(\%)$\\
\hline
Photon reconstruction & $2.0$\\
$\mbc^{\gamma\gamma}$ requirement & $3.1$ \\
ST  $D^0$ yields & $1.0$ \\
\hline
Total &  $3.8$ \\
\hline
\multicolumn{2}{c}{} \\
\multicolumn{2}{c}{Systematic variations due to fitting procedure} \\
\hline
 Source & Variations\\
\hline
Fit range (GeV) & $\pm0.02$ in $E^{\text{tag}}$ and $\pm0.05$ in  $E^{\gamma\gamma}$\\
$D^0\to\pi^0\pi^0$ norm. & $\pm4.6\%$  \\
$D^0\to\pi^0\pi^0$ shape & Smear in $\Delta E^{\text{tag}}$\\
$D^0\bar{D^0}$ bkg shape
              & $\Delta
                \b_\text{input}[D^0\to(\eta\pi^0/\eta\eta/K^0_L\pi^0/K^0_L\eta)]$  \\
Non-$D^0\bar{D}^0$ bkg shape & Flat vs Linear \\
 Signal shape & Smear in $\Delta  E^{\text{tag}}$ and  $\Delta
                E^{\gamma\gamma}$, shift in  $E^{\gamma\gamma}$ \\
\hline
\end{tabular}
}
\end{center}
\end{table}

\section{\boldmath The result for $D^0 \to \gamma \gamma$}\label{sec:dtgamgamresult}

Since we do not observe a 
signal, we set an upper limit
on the branching fraction for $D^0 \to \gamma \gamma$.
We first obtain a smooth background-only PDF shape from the sample
via the kernel estimation method~\cite{keypdf}.
This is done by utilizing the RooFit class~\cite{roofit} RooNDKeysPdf~\cite{roofitkeypdf}.
We then generate $2.2$ million toy MC samples 
by randomly distributing points according to the PDF shape,
while taking into account the Poisson distribution.
We fit to each of these toy samples while randomly making systematic variations in
the fitting procedure, as described in the previous section. We also simultaneously
smear each of the fitted branching fractions with
a Gaussian whose width ($3.8\%$) corresponds to the total systematic uncertainty
that is not associated with the fitting procedure.

\begin{figure}[htb]
\begin{center}
\includegraphics[keepaspectratio=true,width=3.3in,angle=0]{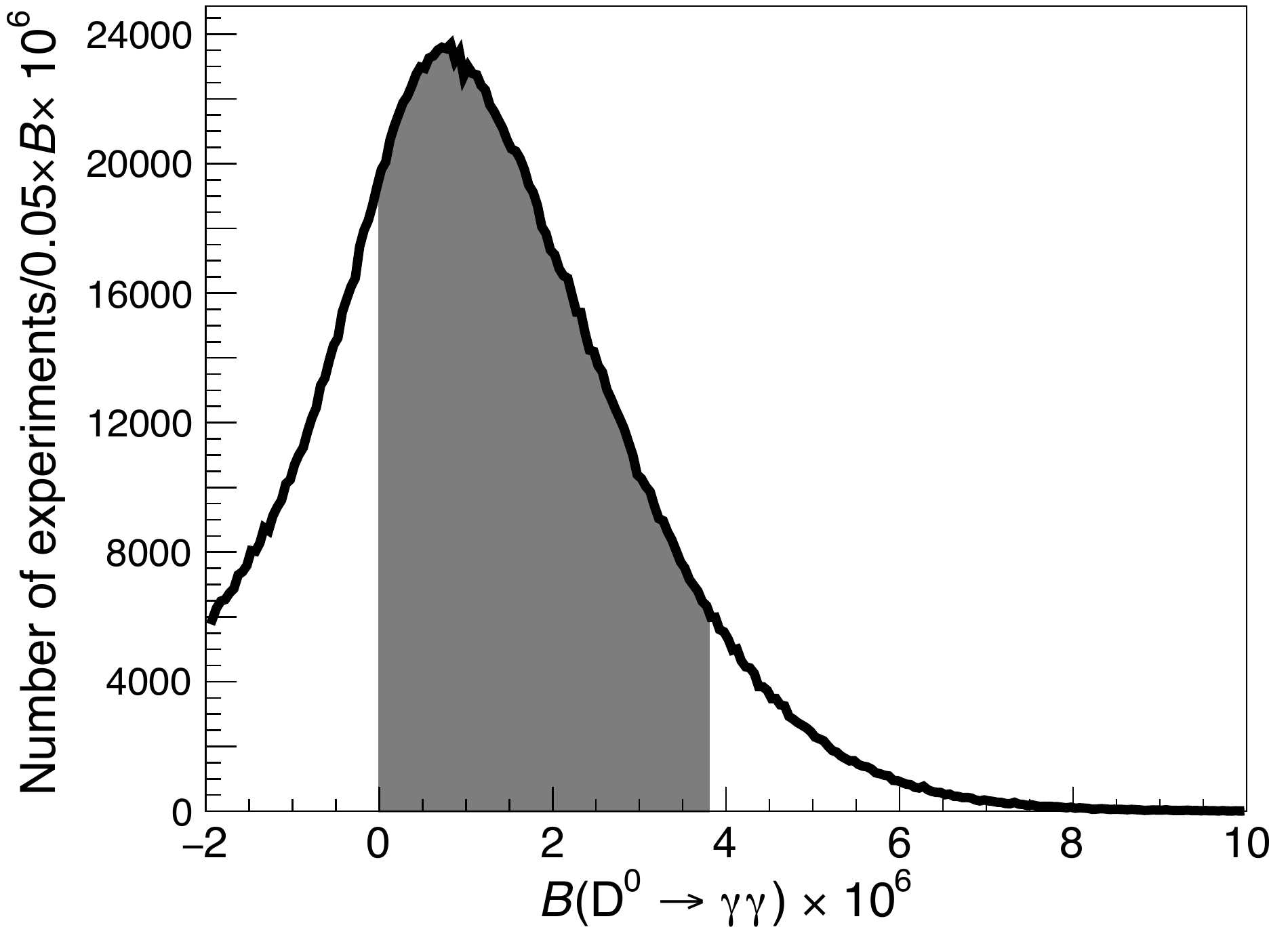}
\caption{Accumulated branching fraction distribution based on
  toy MC samples 
generated from the data-driven PDF.  (See the text for details.)
The shaded region represents $90\%$ of the physical region.
}\label{fig:dtggdtsysttoy}
\end{center}
\end{figure}

Figure~\ref{fig:dtggdtsysttoy} shows an accumulation of the resulting branching
fractions for $D^0 \to \gamma\gamma$. The shaded region represents $90\%$
of its physical region, which we use to set our $90\%$ CL upper limit
of $\b(D^0\to \gamma\gamma) < 3.8 \times 10^{-6}$. If the systematic uncertainty 
were ignored in setting this limit it would be reduced by $0.1\times10^{-6}$.
The expected measurement of branching fraction from these toy experiments is 
 $(+0.7^{+2.0}_{-2.5})\times10^{-6}$, where the quoted uncertainties
 correspond to $68\%$ of the areas under the curves in
 Fig.~\ref{fig:dtggdtsysttoy}. 
The mean value of the accumulated branching fractions  is consistent
with the value of the branching fraction from the nominal fit to data at $0.6\sigma$ level.

%_/_/_/_/_/_/_/_/_/_/_/_/_/_/_/_/_/_/_/_/_/_/_/_/_/_/_/_/_/_/_/_/_/_/_/_/_/_/
\section{\boldmath Improved measurement of $D^0 \to \pi^0\pi^0$ branching fraction }\label{sec:nontag}
 
 As a byproduct of this analysis we also measure the branching fraction of $D^0 \to
 \pi^0 \pi^0$ using the same data sample.
Since the produced $D^0\bar{D}^0$ pairs in our sample necessarily have
opposite $CP$ eigenvalues~\cite{y_cp}, the effective branching fraction for the
$CP$-even final state $\pi^0\pi^0$ is altered when it is measured in events
tagged with a $CP$-mixed state such as $\bar{D}^0\to K^+\pi^-$~\cite{CLEOkpi}.
To avoid this complication and to improve the statistics, 
instead of a DT technique, 
 we reconstruct only one $D^0$ or $\bar{D}^0$ decay in the $\psi(3770) \to D^0\bar{D}^0$ process.  The observed yield is normalized to the 
 total number of  the $D^0\bar{D}^0$ pairs, which can be obtained as 
$N_{D^0\bar{D}^0} = {\cal L} \times \sigma(e^+e^- \to \psi(3770) \to
D^0\bar{D}^0)$,
using the integrated luminosity  ${\cal L}$ of our sample~\cite{bes3lumi} and the previously measured cross section
$ \sigma(e^+e^- \to D^0\bar{D}^0) = (3.607\pm 0.017
(\text{stat.}) \pm0.056  (\text{syst.}))$~nb ~\cite{He:2005bs}.   The branching fraction 
for $D^0 \to \pi^0 \pi^0$ can be calculated as 
\[
\b(D^0\to\pi^0\pi^0) = \frac{N_{\pi^0\pi^0}}{\epsilon_{\pi^0\pi^0}\cdot 2 N_{D^0\bar{D}^0}} , 
\]
where $N_{\pi^0\pi^0}$ is the observed number of $D^0 \to \pi^0 \pi^0$ decays and $\epsilon_{\pi^0\pi^0}$ is the selection efficiency determined with MC. 

The reconstruction of $\pi^0$ candidates is the same as those in the ST modes described in Sec.~\ref{subsec:stag}.
We choose a pair of reconstructed $\pi^0$s that give
the smallest $|\Delta E^{\pi^0\pi^0}|$, and
require $-0.06<\Delta E^{\pi^0\pi^0}<+0.03$~GeV. 
The resolution of $\Delta E^{\pi^0\pi^0}$ is about $20$~MeV.   Then we extract the signal
yield from a fit to  $\mbc^{\pi^0\pi^0}$.  The efficiency is determined to be $\epsilon_{\pi^0\pi^0}=36$\% from MC simulations.

\begin{figure}[htb]
\begin{center}
\includegraphics[keepaspectratio=true,width=3.3in,angle=0]{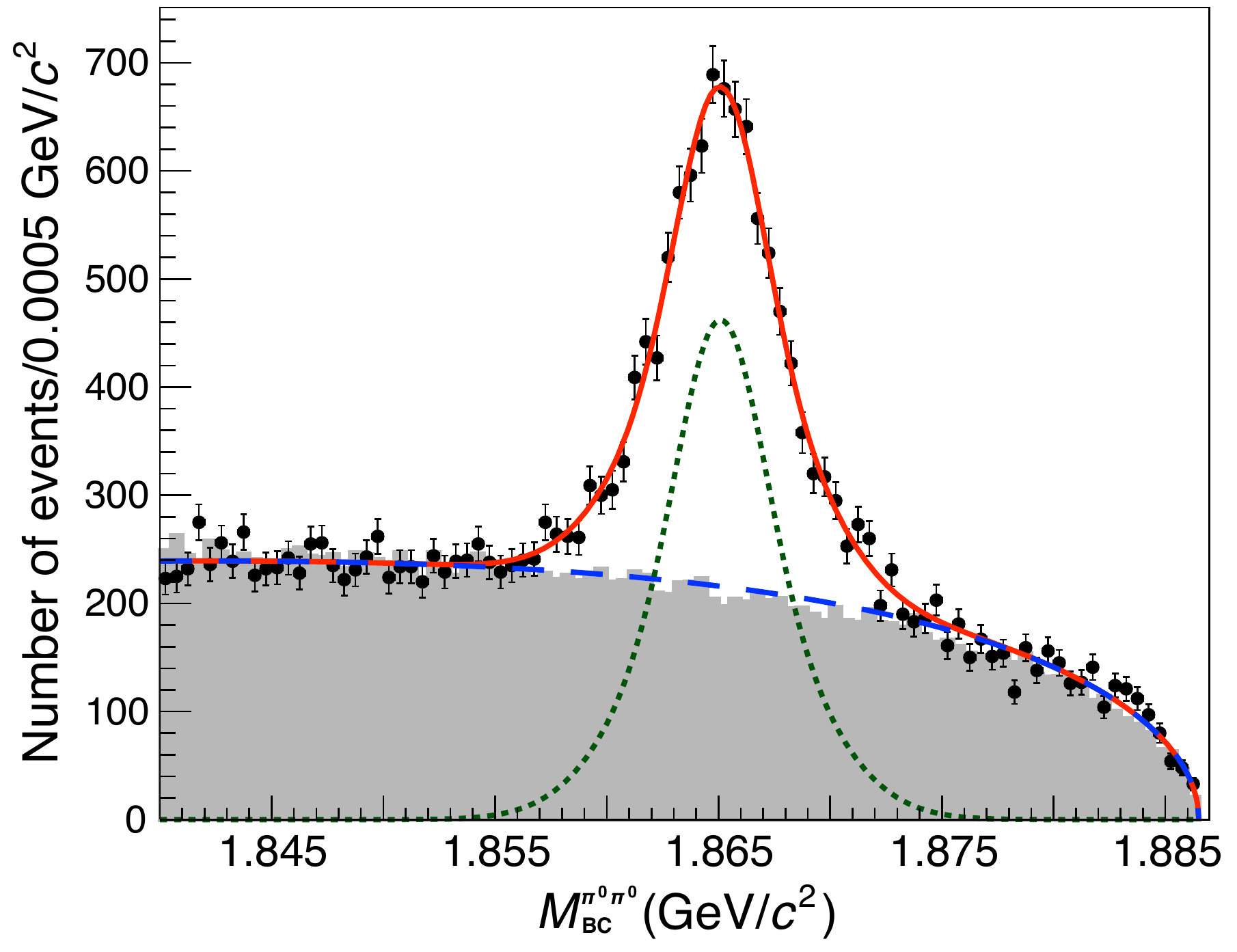}
\caption{Fit to the $\mbc^{\pi^0\pi^0}$ distribution in data for $D^0 \to \pi^0\pi^0$ candidates (points). 
The shaded histogram is the background predicted by MC.  The solid and dashed curves 
are the total fit and the background component, respectively,
and the dotted curve shows the signal. 
}\label{fig:st00dtfit}
\end{center}
\end{figure}

Figure~\ref{fig:st00dtfit} shows a fit to the $\mbc^{\pi^0\pi^0}$
distribution in $1.8400< \mbc^{\pi^0\pi^0}<1.8865$~GeV/$c^2$.
We use a double-Gaussian function to describe the signal shape, which is shown as a dotted line,
and the background shape is described by
an ARGUS background function~\cite{argusBKG}.
From this fit, which yields $\chi^2/$d.o.f. $=91.8/85$, we obtain $N_{\pi^0\pi^0}= 6277\pm156$ events.
In Fig.~\ref{fig:st00dtfit}, we also overlay the backgrounds that
are estimated by the MC simulations (gray shaded histogram).

From the fitted signal yields ($N_{\pi^0\pi^0}$) and reconstruction
efficiency ($\epsilon_{\pi^0\pi^0}$), we obtain
\[
\b(D^0\to\pi^0\pi^0) =  (8.24\pm0.21 (\text{stat.}) \pm0.30 (\text{syst.}))\times10^{-4}.
\]
The quoted total systematic uncertainty ($3.6\%$) is the quadrature sum of
the following seven sources of uncertainty.
(i) The uncertainty due to $\pi^0$ reconstruction is estimated with a DT $D^0\to
K^-\pi^+\pi^0$ sample.
(ii) Histogram binning scheme is varied.
(iii) Narrower ($1.8450<\mbc^{\pi^0\pi^0}<1.8820$~GeV/$c^2$) and
broader  ($1.8350<\mbc^{\pi^0\pi^0}<1.8865$~GeV/$c^2$) fit ranges are tried.
(iv) Narrower ($-0.055<\Delta E^{\pi^0\pi^0}<0.025$~GeV) and broader
($-0.065<\Delta E^{\pi^0\pi^0}<0.035$~GeV) requirements are applied.
(v) Instead of using the ARGUS function~\cite{argusBKG}, a MC-based
background shape is used.
(vi) To assess a posible bias due to the signal line shape,
we fix the all shape parameters of the double Gaussians
 based on the shape extracted from the DT $D^0\to K^-\pi^+\pi^0$ 
sample.
(vii) The uncertainty of the determination of $N_{D^0\bar{D}^0}$ is
determined based on Refs.~\cite{bes3lumi,He:2005bs}.
The resultant relative uncertainties are shown in Table~\ref{tab:nt00syst}.

\begin{table}[htb]
\caption{Systematic uncertainties for $D^0\to\pi^0\pi^0$ analysis.
}
\label{tab:nt00syst}
\begin{center}
\scalebox{1.0}
{
\begin{tabular}{c| c }

\hline
 Source & Relative uncertainty $(\%)$\\
\hline
$\pi^0$ reconstruction & $1.5$ \\
Histogram binning        & $0.1$ \\
Fit range                        & $2.4$ \\
$\Delta E^{\pi^0\pi^0}$ requirement & $0.6$ \\
Background shape         & $0.2$ \\
Signal shape                  & $0.9$ \\
$N_{D^0\bar{D}^0}$            & $1.9$ \\
\hline
Total &  $3.6$ \\
\hline
\end{tabular}
}
\end{center}
\end{table}

\section{\boldmath Conclusions}\label{sec:concl}

Using $2.92$ fb$^{-1}$ of $e^+e^-$ annihilation data collected at
$\sqrt{s} = 3.773$~GeV with the BESIII detector,
we have searched for the FCNC decay $D^0\to\gamma\gamma$ and observe no significant signal.
We set an upper limit $\b(D^0\to\gamma\gamma) < 3.8\times10^{-6}$
at the $90\%$ CL, which is consistent with the upper limit previously
set by the {\it BABAR}~Collaboration~\cite{babar} and with the SM prediction.  
Ours is the first experimental study of this decay using data at
open-charm threshold. Employing the DT technique,  we are able
to suppress the backgrounds from non-$D\bar{D}$ decays effectively.
Our analysis also shows that 
the peaking background from $D^0 \to \pi^0\pi^0$ can be reliably
estimated with a data-driven method.    

We have also measured the branching fraction for $D^0 \to \pi^0\pi^0$
to be  
$(8.24\pm0.21 (\text{stat.}) \pm0.30 (\text{syst.}))\times10^{-4}$
which is consistent with the previous measurements~\cite{pdg14}
and the most precise to date.

\begin{acknowledgments}
The BESIII collaboration thanks the staff of BEPCII and the IHEP
computing center for their strong support. This work is supported in
part by National Key Basic Research Program of China under Contract
No.~2015CB856700; National Natural Science Foundation of China (NSFC)
under Contracts Nos.~11125525, 11235011, 11322544, 11335008, 11425524;
the Chinese Academy of Sciences (CAS) Large-Scale Scientific Facility
Program; Joint Large-Scale Scientific Facility Funds of the NSFC and CAS 
under Contracts Nos.~11179007, 11179014, U1232201, U1332201; CAS under
Contracts Nos.~KJCX2-YW-N29, KJCX2-YW-N45; 100 Talents Program of CAS;
INPAC and Shanghai Key Laboratory for Particle Physics and Cosmology;
German Research Foundation DFG under Contract No.~Collaborative
Research Center CRC-1044; Istituto Nazionale di Fisica Nucleare,
Italy; Ministry of Development of Turkey under Contract
No.~DPT2006K-120470; Russian Foundation for Basic Research under
Contract No.~14-07-91152; U.S.~Department of Energy under Contracts
Nos.~DE-FG02-04ER41291, DE-FG02-05ER41374, DE-FG02-94ER40823,
DESC0010118; U.S.~National Science Foundation; University of Groningen
(RuG) and the Helmholtzzentrum fuer Schwerionenforschung GmbH (GSI),
Darmstadt; WCU Program of National Research Foundation of Korea under
Contract No.~R32-2008-000-10155-0.
\end{acknowledgments}


\begin{thebibliography}{**}
\bibitem{gim1970} S.~L.~Glashow, J.~Iliopoulos, and L.~Maiani,
\href{http://journals.aps.org/prd/abstract/10.1103/PhysRevD.2.1285}
{Phys.\ Rev.\ D {\bf 2}, 1285 (1970)}.

\bibitem{greub1996} C.~Greub, T.~Hurth, M.~Misiak, and D.~Wyler,
\href{http://www.sciencedirect.com/science/article/pii/0370269396006946}
{Phys.\ Lett.\ B {\bf 382}, 415 (1996)}.

\bibitem{fajfer2001} S.~Fajfer, P.~Singer, and J.~Zupan, 
\href{http://journals.aps.org/prd/abstract/10.1103/PhysRevD.64.074008}
{Phys.\ Rev.\ D {\bf 64}, 074008 (2001)}.

\bibitem{burdman2002} G.~Burdman, E.~Golowich, J.~A.~Hewett, and
  S.~Pakvasa, 
\href{http://journals.aps.org/prd/abstract/10.1103/PhysRevD.66.014009}
{Phys.\ Rev.\ D {\bf 66}, 014009 (2002)}.

\bibitem{Prelovsek2001} S.~Prelovsek and D.~Wyler,
\href{http://www.sciencedirect.com/science/article/pii/S0370269301000776}
{Phys.\ Lett.\ B {\bf 500}, 304 (2001)}.

\bibitem{bigi2010} A.~Paul, I.~I.~Bigi, and S.~Recksiegel,
\href{http://journals.aps.org/prd/abstract/10.1103/PhysRevD.82.094006}
{Phys.\ Rev.\ D {\bf 82}, 094006 (2010)}.

\bibitem{babar} J.~P.~Lees {\it et al.} ({\it BABAR} Collaboration),  
\href{http://journals.aps.org/prd/abstract/10.1103/PhysRevD.85.091107}
{Phys.\ Rev.\ D {\bf 85}, 091107(R) (2012)}.


\bibitem{cleo} T.~E.~Coan {\it et al.} (CLEO Collaboration),
\href{http://journals.aps.org/prl/abstract/10.1103/PhysRevLett.90.101801}
{Phys.\ Rev.\ Lett.\ {\bf 90}, 101801 (2003)}.


\bibitem{bes3lumi} M.~Ablikim {\it et al.} (BESIII Collaboration), 
\href{http://iopscience.iop.org/1674-1137/37/12/123001/pdf/1674-1137_37_12_123001.pdf}
{Chin.\ Phys.\ C {\bf 37}, 123001 (2013)}.


\bibitem{He:2005bs} 
  Q.~He {\it et al.} (CLEO Collaboration),
  \href{http://journals.aps.org/prl/abstract/10.1103/PhysRevLett.95.121801}
{Phys.\ Rev.\ Lett.\  {\bf 95}, 121801 (2005)};
  \href{http://journals.aps.org/prl/abstract/10.1103/PhysRevLett.96.199903}
  {{\bf 96}, 199903(E) (2006)};
  S.~Dobbs {\it et al.}  (CLEO Collaboration),
\href{http://journals.aps.org/prd/abstract/10.1103/PhysRevD.76.112001}
{Phys.\ Rev.\ D {\bf 76}, 112001 (2007)};
  G.~Bonvicini {\it et al.}  (CLEO Collaboration),
\href{http://journals.aps.org/prd/abstract/10.1103/PhysRevD.89.072002}
{Phys.\ Rev.\ D {\bf 89}, 072002 (2014)};
   \href{http://journals.aps.org/prd/abstract/10.1103/PhysRevD.91.019903}
  {{\bf 91}, 019903(E) (2015)}.

\bibitem{marck3} R.~M.~Baltrusaitis {\it et al.} (MARK III Collaboration),
\href{http://journals.aps.org/prl/abstract/10.1103/PhysRevLett.56.2140}
{Phys.\ Rev.\ Lett.\ {\bf 56}, 2140 (1986)}.

\bibitem{Grossman:2012ry} 
  W.~Kwong and S.~P.~Rosen,
\href{http://www.sciencedirect.com/science/article/pii/037026939391843C}
{Phys.\ Lett.\ B {\bf 298}, 413 (1993)};
  Y.~Grossman and D.~J.~Robinson,
\href{http://link.springer.com/article/10.1007/JHEP04%282013%29067}
{J. High Energy Phys. 04 ({\bf 2013}) 067}.

\bibitem{Grossman:2012eb} 
  Y.~Grossman, A.~L.~Kagan and J.~Zupan,
\href{http://journals.aps.org/prd/abstract/10.1103/PhysRevD.85.114036}
{Phys.\ Rev.\ D {\bf 85}, 114036 (2012)}.
  
\bibitem{bes3det} M.~Ablikim {\it et al.} (BESIII Collaboration),
\href{http://www.sciencedirect.com/science/article/pii/S0168900209023870}
{Nucl.\ Instrum.\ Methods\ Phys.\ Res., \ Sec.\ A {\bf 614}, 345 (2010)}.


\bibitem{kkmc} S.~Jadach, B.~F.~L.~Ward and Z.~Was,
\href{http://journals.aps.org/prd/abstract/10.1103/PhysRevD.63.113009}
{Phys.\ Rev.\ D {\bf 63}, 113009 (2001)}.

\bibitem{evtgen} D.~J.~Lange, 
\href{http://www.sciencedirect.com/science/article/pii/S0168900201000894}
{Nucl.\ Instrum.\ Methods Phys.\ Res., Sect.\ A {\bf 462}, 152
  (2001)}; R.~G.~Ping,
\href{http://iopscience.iop.org/1674-1137/32/8/001}
{Chin.\ Phys.\ A {\bf 32}, 599 (2008)}.

\bibitem{pdg} K.~Nakamura {\it et al.} (Particle Data Group), 
\href{http://pdg.lbl.gov/2011/listings/contents_listings.html}
{J.\ Phys.\ G {\bf 37}, 075021 (2010) and 2011 partial update for the 2012
  edition}.

\bibitem{babayaga} 
   C.M.~Carloni Calame {\it et al.},
\href{http://www.sciencedirect.com/science/article/pii/S0550321300003564}
{Nucl.\ Phys.\ {\bf B584}, 459(2000)};
   C.M.~Carloni Calame {\it et al.}, 
\href{http://www.sciencedirect.com/science/article/pii/S037026930101108X}
{Phys.\ Lett.\ B {\bf 520}, 16 (2001)};
   C.M.~Carloni Calame {\it et al.}, 
   \href{http://www.sciencedirect.com/science/article/pii/S0920563204000106}
   {Nucl.\ Phys.\ B, Proc.\ Suppl.\ {\bf 131}, 48 (2004)};
   G.~Balossini {\it et al.}, 
   \href{http://www.sciencedirect.com/science/article/pii/S0550321306007851}
   {Nucl.\ Phys.\ {\bf B758}, 227 (2006)};
   G.~Balossini, {\it et al.}, 
\href{http://www.sciencedirect.com/science/article/pii/S0370269308004346}
{Phys.\ Lett.\ B {\bf 663}, 209 (2008)};

\bibitem{geant}  S.~Agostinelli {\it et al.},
\href{http://www.sciencedirect.com/science/article/pii/S0168900203013688}
{Nucl.\ Instrum.\ Methods Phys.\ Res., Sec.\ A {\bf 506}, 250 (2003)};
 J.~Allison {\it et al.},
\href{http://ieeexplore.ieee.org/xpls/abs_all.jsp?arnumber=1610988&tag=1}
{IEEE\ Trans.\ Nucl.\ Sci.\ {\bf 53}, 270 (2006)};
Z.~Y.~Deng  {\it et al.},
\href{http://hepnp.ihep.ac.cn/qikan/epaper/zhaiyao.asp?bsid=5105}
{High\ Energy\ Physics\ and\ Nuclear\ Physics {\bf 30}, 371 (2006)}.

\bibitem{y_cp} M.~Ablikim {\it et al.} (BESIII Collaboration),  
\href{http://www.sciencedirect.com/science/article/pii/S0370269315002518}{Phys.\ Lett.\ B {\bf 744}, 339 (2015)}.

\bibitem{argusBKG} H.~Albrecht {\it et al.} (ARGUS Collaboration),  
\href{http://www.sciencedirect.com/science/article/pii/037026939091293K}
{Phys.\ Lett.\ B {\bf 241}, 278 (1990)}.

\bibitem{cblshape} J.~ E.~Gaiser, Ph.D.~thesis,
Stanford Linear Accelerator Center, Stanford University, Stanford,
California [Report No.
\href{http://www.slac.stanford.edu/pubs/slacreports/slac-r-255.html}
{SLAC-R-255, 1982} (unpublished)]; 
T. Skwarnicki, Ph.D.~thesis, Institute of Nuclear Physics, Krakow, Poland
[Report No.
\href{http://inspirehep.net/record/230779/files/f31-86-02.pdf}
{DESY-F31-86-02, 1986} (unpublished)].

\bibitem{keypdf} Kyle~S.~Cranmer,
\href{http://www.sciencedirect.com/science/article/pii/S0010465500002435} 
{Comput.\ Phys.\ Commun.\ {\bf 136}, 198 (2001)}.

\bibitem{roofit} W.~Verkerke and D.~Kirkby, 
\href{http://www.slac.stanford.edu/econf/C0303241/proc/papers/MOLT007.PDF} 
{eConf No.~C0303241 (2003) MOLT007}
[\href{http://arxiv.org/abs/physics/0306116} 
{arXiv:physics/0306116}].

\bibitem{roofitkeypdf} 
\href{https://root.cern.ch/root/html/RooNDKeysPdf.html} 
{https://root.cern.ch/root/html/RooNDKeysPdf.html}.

\bibitem{CLEOkpi} D.~M.~Asner {\it et al.} (CLEO Collaboration), 
\href{http://journals.aps.org/prd/abstract/10.1103/PhysRevD.86.112001}
{Phys.\ Rev.\ D {\bf 86}, 112001 (2012)}.

\bibitem{pdg14} K.~A.~Olive {\it et al.} (Particle Data Group), 
\href{http://pdg.lbl.gov/2014/listings/contents_listings.html}
{Chin.\ Phys.\ C {\bf 37}, 090001 (2014)}.

\end{thebibliography}
\end{document}